\documentclass[acmsmall, nonacm, screen]{acmart}

\usepackage{tcolorbox}
\usepackage{xspace}
\usepackage{enumitem}
\usepackage{caption}
\usepackage{subcaption}
\usepackage{bbding}
\usepackage{url}
\usepackage{booktabs}
\usepackage{tabularx}
\usepackage{circledsteps} 
\usepackage{graphicx}
\usepackage{footmisc}

\usepackage{xcolor}
\usepackage{caption}
\usepackage{listings}
\definecolor{mygreen}{rgb}{0,0.6,0}
\definecolor{mymauve}{rgb}{0.58,0,0.82}
\lstdefinelanguage{JavaScript}{
  keywords={typeof, new, true, false, catch, function, return, null, catch, switch, var, if, in, while, do, else, case, break, const, let, async, await, for, of},
  ndkeywords={class, export, boolean, throw, implements, import, this, console, log},
  sensitive=false,
  comment=[l]{//},
  morecomment=[s]{/*}{*/},
  morestring=[b]",
}
\lstdefinelanguage{Rust}{
    keywords={fn, let, mut, pub, struct, enum, impl, trait, where, match, if, else, for, in, while, loop, break, continue, return, async, await, move, dyn, const, static, ref, use, mod, crate, super, self, Self, as, extern, type, true, false, Option, Some, None, Result, Ok, Err, Box, Vec, String, usize, i32, u32, f64, bool, char, str},
    ndkeywords={println!, vec!, macro_rules!, assert_eq!, panic!},
    sensitive=true,
    comment=[l]{//},
    morecomment=[s]{/*}{*/},
    morestring=[b]',
    morestring=[b]",
}
\usepackage[capitalise]{cleveref}
\usepackage{pifont}

\usepackage{multicol}

\usepackage{makecell}
\usepackage{multirow}
\newcolumntype{C}[1]{>{\centering\arraybackslash}p{#1}}

\usepackage{amsmath,bm,amsthm}
\theoremstyle{definition}

\usepackage{algorithm}
\usepackage{algpseudocode}
\algnewcommand\algorithmicforeach{\textbf{for each}}
\algdef{SE}[FUNCTION]{Function}{EndFunction}[2]{\algorithmicfunction\ \textnormal{#1}\ (#2)}{\algorithmicend\ \algorithmicfunction}%
\algdef{S}[FOR]{ForEach}[1]{\algorithmicforeach\ #1\ \algorithmicdo}
\algrenewcommand\alglinenumber[1]{\footnotesize #1}
\algrenewcommand\algorithmicrequire{\small \textbf{input:}}
\algrenewcommand\algorithmicensure{\small \textbf{output:}}
\algrenewcommand\algorithmicfunction{\textbf{Function}}
\algtext*{EndFunction} %
\algtext*{EndIf} %
\algtext*{EndFor} %
\algtext*{EndWhile} %

\usepackage{fontawesome} 
\usepackage{xcolor}
\usepackage{framed}
\usepackage{textcomp}
\usepackage{adjustbox}

\usepackage{microtype}
\usepackage{ifthen}

\def\editmode{} 

\ifthenelse{\isundefined{\editmode}}{
    \newcommand{\editnote}[3]{%
    }
}
{
    \newcommand{\editnote}[3]{\xspace\colorbox{#1}{\sffamily \smaller \textcolor{white}{~\faCommenting{}~#2~}}\textcolor{#1}{~#3}\xspace}
}
\definecolor{latte-teal}{RGB}{23, 146, 153}
\definecolor{nord0}{HTML}{2E3440}
\definecolor{nord1}{HTML}{3B4252}
\definecolor{nord2}{HTML}{434C5E}
\definecolor{nord3}{HTML}{4C566A}
\definecolor{nord4}{HTML}{D8DEE9}
\definecolor{nord5}{HTML}{E5E9F0}
\definecolor{nord6}{HTML}{ECEFF4}
\definecolor{nord7}{HTML}{8FBCBB}
\definecolor{nord8}{HTML}{88C0D0}
\definecolor{nord9}{HTML}{81A1C1}
\definecolor{nord10}{HTML}{5E81AC}
\definecolor{nord11}{HTML}{BF616A}
\definecolor{nord12}{HTML}{D08770}
\definecolor{nord13}{HTML}{EBCB8B}
\definecolor{nord14}{HTML}{A3BE8C}
\definecolor{nord15}{HTML}{B48EAD}

\begin{document}

\title{Multi-Agent Systems for Dataset Adaptation in Software Engineering: Capabilities, Limitations, and Future Directions}

\author{Jingyi Chen}
\authornote{Co-first authors.}
\email{jchenix@connect.ust.hk}
\affiliation{%
    \institution{The Hong Kong University of Science and Technology}
    \city{Hong Kong}\country{China}
}

\author{Xiaoyan Guo}
\authornotemark[1]
\email{floraa@mail.ustc.edu.cn}
\affiliation{%
    \institution{University of Science and Technology of China}
    \city{Hefei}\country{China}
}

\author{Songqiang Chen}
\email{i9s.chen@connect.ust.hk}
\affiliation{%
    \institution{The Hong Kong University of Science and Technology}
    \city{Hong Kong}\country{China}
}

\author{Shing-Chi Cheung}
\authornote{Corresponding authors.}
\email{scc@cse.ust.hk}
\affiliation{%
    \institution{The Hong Kong University of Science and Technology}
    \city{Hong Kong}\country{China}
}

\author{Jiasi Shen}
\authornotemark[2]
\email{sjs@cse.ust.hk}
\affiliation{%
    \institution{The Hong Kong University of Science and Technology}
    \city{Hong Kong}\country{China}
}

\begin{abstract}
Automating the adaptation of software engineering (SE) research artifacts across datasets is essential for scalability and reproducibility, yet it remains largely unstudied. Recent advances in large language model (LLM)–based multi-agent systems, such as GitHub Copilot’s agent mode, promise to automate complex development workflows through coordinated reasoning, code generation, and tool interaction. This paper presents the first empirical study on how state-of-the-art multi-agent systems perform in dataset adaptation tasks. We evaluate Copilot, backed by GPT-4.1 and Claude Sonnet 4, on adapting SE research artifacts from benchmark repositories including ROCODE and LogHub2.0. Through a five-stage evaluation pipeline (file comprehension, code editing, command generation, validation, and final execution), we measure success rates, analyze failure patterns, and assess prompt-based interventions designed to enhance agent performance. Results show that current systems can identify key files and generate partial adaptations but rarely produce functionally correct implementations. 
Prompt-level interventions, especially providing execution error messages and reference code, substantially improve structural similarity to ground truth (from 7.25\% to 67.14\%), highlighting the importance of contextual and feedback-driven guidance. Our findings reveal both the promise and limitations of today’s multi-agent LLM systems for dataset adaptation, and suggest concrete directions for building more reliable, self-correcting agents in future SE research.
\end{abstract}


\maketitle

\section{Introduction}

Software Engineering (SE) has entered a period of rapid innovation. In areas ranging from code generation and bug fixing to performance analysis and security, researchers are constantly proposing new techniques\cite{Requirements-Driven, Decade-of-Progress, survey-on-dsl}. Each technique must be evaluated on multiple datasets to demonstrate its general effectiveness beyond a single repository or environment \cite{Loghub-2.0, ClassEval}; yet automatic adaptation across datasets remains largely unexplored. Human experts can manually tailor a technique to each dataset (e.g., installing dependencies, building pipelines, and crafting test cases), but this approach does not scale as the number and diversity of datasets grow. The need for scalable, automated dataset adaptation is therefore critical for future SE research.

Large language models open a new possibility. Modern LLMs, such as ChatGPT and GitHub Copilot, can interpret code, search project directories, generate commands, modify files, and interact with external tools. These capabilities can be orchestrated in multi‑agent \cite{TOSEM-LLM-based-agent4SE} systems, where multiple specialized LLM agents collaborate to plan and execute complex tasks. Recent surveys show that multi‑agent systems leverage cross‑examination and role specialization to provide autonomous problem solving, robustness, and scalability across software‑development activities. GitHub Copilot's agent mode exemplifies the multi-agent systems: the agent analyzes a repository, plans multi‑step solutions, runs commands or tests, uses MCP tools \cite{MCP} to browse repositories and edit files, and iteratively refines its work. When combined with emerging benchmarks, multi-agent systems could autonomously adapt software artifacts: parsing repository metadata, generating build commands, correcting errors, and verifying outcomes, thus reducing human effort and enabling reproducibility at scale.

However, the effectiveness of LLM‑based multi‑agent systems in automating dataset adaptation remains unknown. There is no empirical study of the performance of state-of-the-art (SOTA) agents when tasked with adapting technologies to new datasets. Understanding their limitations is crucial for future work to design reliable multi-agent systems for dataset adaptation tasks. This paper presents the first empirical study to evaluate multi‑agent systems on this challenging task. We investigate how a SOTA agent, represented by GitHub Copilot's agent mode, adapts diverse SE tools to different benchmark datasets. Our two research questions guide the study:

\textbf{RQ1:} \textit{How does a SOTA multi‑agent system perform when automatically adapting software engineering research artifacts to new datasets? }We detail the steps taken by the agent, measure success rates, and identify failure modes.

\textbf{RQ2: } \textit{Can prompt‑level interventions improve performance and inform the design of more effective multi‑agent systems for this task?} Inspired by recent work on prompt engineering and agent coordination~\cite{SWE-agent}, we explore interventions such as providing error messages, injecting missing information, or prompting fault locations to instruct agents to retry upon failure.

By answering these questions, we aim to illuminate both the promise and limitations of LLM‑based multi‑agent systems for automated dataset adaptation. Our findings will inform the design of next‑generation agents and contribute to more scalable software engineering research.

\section{Problem Formulation}

Let $R_D$ denote a dataset repository and $R_T$ denote a technology repository, each provided via a GitHub URL or a local path. Our objective is to automatically modify $R_D$ or $R_T$, construct a runnable experiment, and obtain its execution results. Our problem can be represented by a mapping:
$$f:\ (R_D, R_T) \mapsto E$$
In the mapping, $E$ is the resulting experiment state, including the code adaptations performed to ensure compatibility, the executable scripts or commands derived for running the experiment, and the results produced by executing those scripts or commands to completion. This formulation captures the overall objective of automating the process of integrating a research artifact with a given dataset and obtaining valid execution results.
\section{Study}

\subsection{Data Collection} \label{data-collection}

To construct a representative set of high-quality SE research artifacts, we followed a systematic multi-stage selection process:

\textbf{Artifact Identification.} We began by collecting all papers accepted by the top four SE  (FSE, ICSE, ASE and ISSTA) in 2024-2025 that were either (i) awarded a Reusable Artifact badge or (ii) explicitly identified by the authors as providing reusable artifacts. These two rules ensure that the selected artifacts meet basic standards of availability and reusability.

\textbf{Language Filtering.} To ensure a consistent and manageable experimental environment, we retained only artifacts implemented in Python. This choice allows for reliable environment deployment and isolates our investigation from environment setup challenges, an issue already extensively examined in previous work \cite{name-it}.

\textbf{Scope of Evaluation.} We further filtered the artifacts to include only those that (i) evaluate a single technique across at least two datasets, or (ii) evaluate two or more state-of-the-art (SOTA) techniques on a newly proposed dataset. This criterion ensures that the selected artifacts include enough components that are suitable for our RQs.

Finally, two domain experts independently examined the selected artifacts and their accompanying papers. Two projects (\textit{i.e.,} ROCODE \cite{ROCODE} and LogHub2.0 \cite{Loghub-2.0}) are selected for the following experiments.

\subsection{Data Preparation} \label{data-preparation}
The repositories of the artifacts collected in Section~\ref{data-collection} each contain two or more datasets that have been adapted to the corresponding methods by their authors. For each repository, we select one adaptation as the target task, whose original implementation serves as the ground truth of the dataset adaptation task. To construct the adaptation task, we first integrate the target dataset repository with the corresponding method's repository by either (1) copying the dataset repository $R_D$ into the technology repository $R_T’s$ directory, or (2) copying the technology repository $R_T$ into the dataset repository $R_D’s$ directory. 
The paths of both repositories are explicitly specified in the prompts to ensure the agent can locate and modify the relevant files.

We then remove the implementation corresponding to the target dataset while keeping the other adaptations intact.
Specifically, if multiple dataset adaptations are implemented in standalone files, they can be explicitly detected and removed; if multiple adaptations share common configuration files, we remove only the entries or options specific to the target dataset. This guarantees that the repository remains runnable for the remaining datasets but lacks the code for the dataset adaptations. Therefore, the processed repositories simulate a realistic adaptation scenario.

Then, multi-agent systems are expected to generate the implementation that we removed. The removed implementation is preserved separately as the ground truth, which serves to evaluate whether the code generated or modified by the agents can correctly reproduce the original dataset adaptation. In addition, files associated with other dataset adaptations will be retained as reference code in each repository. These code files will later be used in RQ1 Step 1 (Reading required files) and RQ2 Prompt 2 (Providing reference code).

\subsection{Study Design}
\label{design_RQ1}
\subsubsection{RQ1: How does a SOTA Multi-Agent System Perform?}

First, we aim to evaluate the performance of SOTA multi-agent systems at each stage of the process. 
Each multi-agent system is provided with a processed repository prepared as described in Section~\ref{data-preparation}, along with a simple prompt that specifies the adaptation task, the locations of both $R_D$ and $R_T$, and other necessary information. Once a system has completely finished running, we record its completion status.

Our preliminary experiments indicate that most of the evaluated multi-agent systems failed to complete the assigned tasks. Under such circumstances, nearly all results would be classified as failures, and thus, analyzing end-to-end performance alone is less informative.
To obtain a more concrete understanding of each agent's behavior throughout the entire process, we manually decompose the overall task into five distinct steps.

\textbf{1). Read required files.} Before modifying code or generating commands for a repository, a multi-agent system must browse the essential files in a repository to understand its architecture. To evaluate the multi-agent system's ability to identify these essential files, two human developers attempted the dataset adaptation task and recorded the files they accessed. These files are considered {necessary files}. Then, we tracked the MCP usage records of the multi-agent system to determine which files were read by the agents and compared this list with the set of necessary files. The system is considered to have completed Step 1 only if it reads all the necessary files at the beginning.

\textbf{2). Edit and create necessary files.} When having an overall understanding of the repository, the multi-agent system should modify the repository by creating or editing the files in it, which is the core step of the dataset adaptation task. We identify the required files by comparing them with the ground truth version described in Section \ref{data-preparation}. To evaluate the multi-agent system's performance in a fine-grained manner, we divide this step into two sub-steps. First, the system must edit existing necessary files and create any missing ones. We assess this sub-step by tracking the system's MCP usage records and determining which files were edited or created. Second, the code files created or edited by the multi-agent system must be functionally consistent with the ground truth version. 
We evaluate this sub-step by setting multiple checkpoints within the code files. At each checkpoint, we print selected intermediate outputs to inspect variable values and compare them with the ground truth.
Considering that generating fully correct code for all necessary files is highly challenging, we rate the multi-agent system as a “partial pass (P)” if it completes the first sub-step and produces correct code for some, but not all, necessary files.

\textbf{3). Generate and execute the commands.} After modifying code files, the multi-agent system should generate scripts (\textit{e.g.,} bash files) or commands to implement the pipeline that evaluates the technology in the dataset. The multi-agent system should also automatically execute the scripts or commands by interacting with the terminal. We also divide this step into two sub-steps. The first sub-step evaluates whether generated scripts or commands are equivalent to the ground truth. For example, if the system generates a script named \textit{run\_humaneval.sh}, we examine whether the script includes an equivalent sequence of commands that execute the prediction pipeline and subsequently trigger the evaluation procedure, as in the ground-truth bash file. The second sub-step measures the multi-agent system's ability to execute the scripts or commands automatically. Specifically, a case is considered successful if the system is able to autonomously execute its own generated command, regardless of whether the command itself is correct (i.e., whether sub-step 1 is satisfied).

\textbf{4). Validating and repairing.}
Since it is generally difficult for multi-agents to generate fully correct code in one round, this step focuses on the multi-agent system’s ability to validate intermediate and final outputs, identifying and correcting issues in the code produced during Step 2.
We further divide this step into two sub-steps. First, the multi-agent system needs to recognize the existence of bugs, typically by designing targeted test inputs for critical modules and running small-scale tests on a subset of the dataset. These tests help verify whether the outputs align with expectations and detect anomalies or inconsistencies between the generated results and the expected behavior.
Once any issue is successfully detected by the multi-agent system, the first sub-step is considered completed. Second, the system initiates a repair process, either by automatically revising the faulty code or by recalling the editing or creation step with additional guidance (e.g., error messages or specific instructions). Similarly, once at least one identified issue has been successfully fixed, the second sub-step is regarded completed. 
Through iterative validation and repair, the system is expected to gradually converge and generate a correct and executable implementation.

\textbf{5). Final results.}  
At last, we perform a final comparison between the generated implementation and the ground truth. Specifically, we execute both versions under identical settings and examine their outputs. 
To conduct a rigorous experiment, we exclude non-functional factors such as execution time or memory usage from the evaluation, focusing solely on functional equivalence. If the generated result matches the ground truth in all expected behaviors, this step is marked as successful; otherwise, it is considered a failure. This binary outcome (true or false) serves as the final evaluation metric for determining whether the entire generation process has produced a correct and functionally equivalent implementation.

It is also worth noting that, during task execution, the agents do not always follow a strictly linear sequence of Step 1–5; rather, they may revisit or interleave steps depending on their internal coordination and reasoning process.

\subsubsection{RQ2: Can Prompt‑level Interventions Fix the Failures Identified in RQ1?}
\label{design_RQ2}
The dataset adaption tasks are challenging for multi-agent systems. A straightforward method to improve agents’ performance is through manual intervention. Therefore, we designed a series of fixing prompts. These prompts are intended to guide the agents in identifying, analyzing, and resolving issues that arise during the task execution process, thereby improving their overall self-correction capability.

\textbf{Prompt 1: Providing error messages.} In this setting, the agent receives raw error outputs copied directly from the terminal (\textit{e.g.}, Python tracebacks), which specify the type and location of each failure. These messages serve as explicit diagnostic signals, enabling the model to infer the cause of errors and make targeted revisions. Prior studies \cite{intercode, rlef, perfcodegen} have shown that exposing execution feedback or runtime errors can significantly improve model debugging and repair performance.

\begin{figure}
    \centering
    \includegraphics[width=0.86\linewidth]{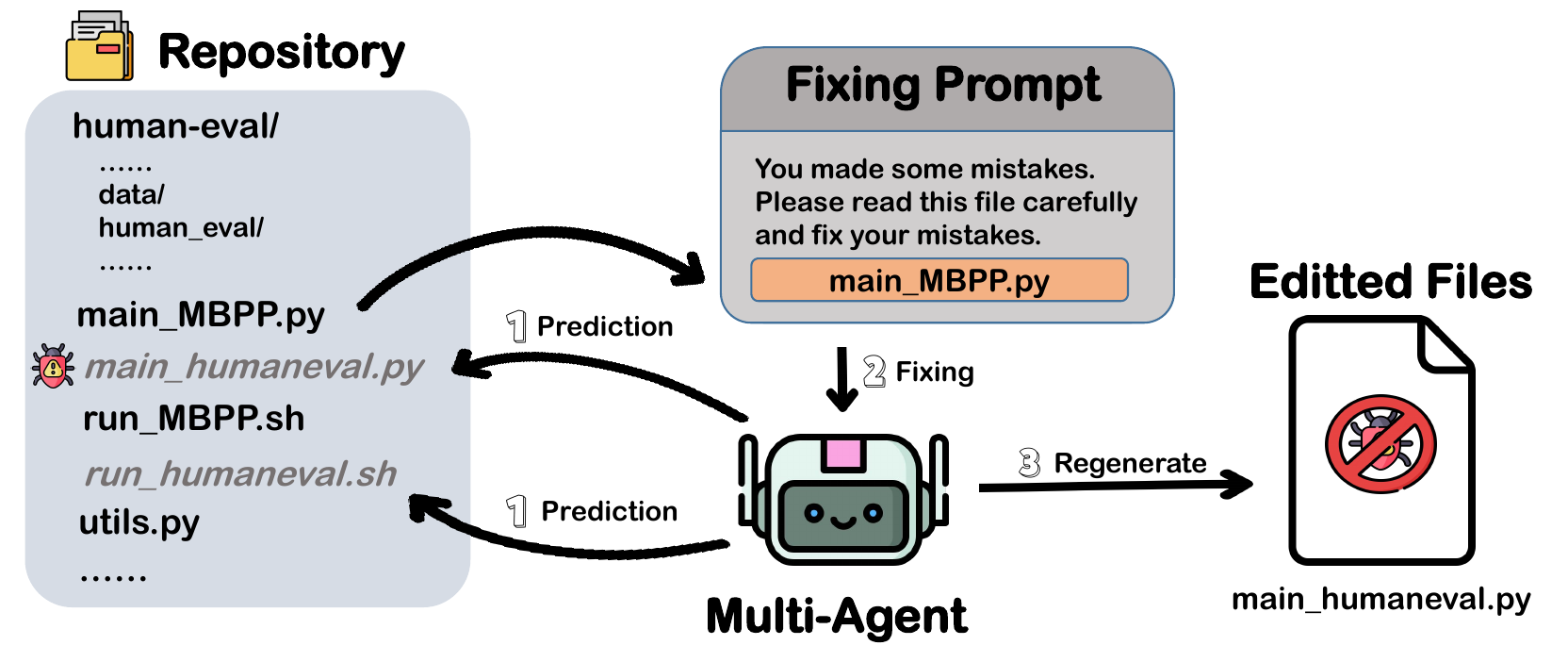}
    \caption{Overall process of constructing and using Prompt 2 in RQ2.}
    \label{fig:RQ2_prompt2}
\end{figure}

\textbf{Prompt 2: Providing reference code.} 
Existing work \cite{codex, repocoder} shows that supplying high-level contextual information, which conveys the structural and semantic relationships among different components in a repository, can improve code adaptation and ensure consistency across environments.
To provide such contextual information, we include the Bash scripts and Python files associated with other dataset adaptations that remain in each repository (as described in Section ~\ref{data-preparation}) in the prompt. 
The overall process of using these reference files from each repository is illustrated in Figure~\ref{fig:RQ2_prompt2}. These reference files provide hints about function usage, parameter formats, and workflow organization, allowing the agent to align its implementation generated in Step~2 with the existing adaptations. When identifying inconsistencies between the reference code and the generated adaptations, Copilot can locate and repair potential issues.
By implicitly revealing where issues might occur and how they could be corrected, the reference code serves as a guiding signal for refining the implementation.

\textbf{Prompt 3: Indicating locations of bugs.} In this setting, an expert manually inspects the intermediate checkpoints and compares them with the ground truth. Once a discrepancy in the intermediate results is identified, the expert traces back along the generation path to locate the earliest erroneous segment, which is then explicitly highlighted in the prompt. This design guides the agent to focus its debugging efforts on the relevant code regions rather than regenerating the entire file from scratch. Similar fine-grained supervision strategies have been reported to improve code correctness \cite{prompt_fix, prompt_pattern} in LLM-based systems. Beyond this general benefit, Prompt 3 in our setup plays a special role as a complementary strategy to Prompt 2. This complementary function manifests in two major aspects: First, based on our observations from RQ1, the agents sometimes fall into a dilemma where Prompt 2 keeps modifying irrelevant parts of the code without fixing the actual issue. In such cases, Prompt 3 helps the agent break out of this loop by explicitly pointing to the exact line that requires modification, guiding it toward the true source of the error. Second, it extends the applicability of the overall setup to cases where reference code is unavailable, thus covering the limitation of Prompt 2. For example, in the MBPP dataset, the function-name extraction step has no counterpart in other datasets, and Prompt 3 allows the agent to handle errors at this stage when Prompt 2 cannot be used.

During the experiments, we generally applied the fixing prompts in the order of Prompt 1, Prompt 2, and Prompt 3. Specifically, Prompt 1 was first used to address syntax or runtime errors. Once the program could execute successfully, Prompt 2 was employed to provide reference code and guide the multi-agent system in correcting logical or structural mistakes. 
If Prompt 2 was applied to the same code issue three consecutive times and the agent either failed to correct the error or repeatedly modified irrelevant code, we then proceeded to Prompt 3. If Prompt 3 demonstrated the same pattern of failure after three consecutive attempts, the issue was deemed unfixable.
However, when the use of Prompt 2 or Prompt 3 resulted in new runtime errors, we reverted to Prompt 1 to restore the executability before continuing with error correction using Prompt 2 and Prompt 3.

\subsection{Environment setup}
We manually configured the environments for all experimental artifacts, as we do not aim to evaluate multi-agent system's performance in environment setup.
Each artifact was set up according to its provided instructions, using Conda environments for dependency management and Docker containers for isolation when applicable.
This manual setup ensured that all artifacts were correctly configured and would be run in a good environment in the following experiments.

After preparing the environments, we proceeded to the dataset adaptation experiments using GitHub Copilot as the representative LLM-based multi-agent system. It was chosen for its advanced code generation capabilities and relevance to real-world development workflows. We evaluated Copilot under two backend LLMs to enhance the robustness of our findings: one backed by OpenAI’s GPT‑4.1 model and another by Anthropic’s Claude Sonnet 4. These backends represent SOTA LLMs, enabling us to assess Copilot’s performance with different underlying AI engines.

\subsection{Results and Analysis}
\subsubsection{RQ1}
\begin{table}[t]
\caption{Results of RQ1. "yes"/"no" means whether the agent successfully complete the Step. "partial" means partially complete (refer to \Cref{design_RQ1}).\label{table:RQ1-steps}}
\begin{adjustbox}{max width=\textwidth}
\begin{tabular}{l|cl|c|c|c|c|c|c|c|c}

\toprule
\multicolumn{1}{c|}{\multirow{2}{*}{\textbf{Agent}}}     & \multicolumn{1}{c|}{\multirow{2}{*}{\textbf{LLM}}}  & \multicolumn{1}{c|}{\multirow{2}{*}{\textbf{Dataset}}} & \multicolumn{1}{c|}{\multirow{2}{*}{\textbf{Step 1}}} & \multicolumn{2}{c|}{\textbf{Step 2}}                             & \multicolumn{2}{c|}{\textbf{Step 3}}                                             & \multicolumn{2}{c|}{\textbf{Step 4}}                               & \multicolumn{1}{c}{\multirow{2}{*}{\textbf{Step 5}}} \\ 
                                               & \multicolumn{1}{l|}{}                         & \multicolumn{1}{l|}{} & \multicolumn{1}{c|}{}             & \multicolumn{1}{c|}{\textbf{Edit}} & \multicolumn{1}{c|}{\textbf{Correct}} & \multicolumn{1}{c|}{\textbf{Generate}} & \multicolumn{1}{c|}{\textbf{Execute}} & \multicolumn{1}{c|}{\textbf{Notice}} & \multicolumn{1}{c|}{\textbf{Fix}} & \multicolumn{1}{c}{}                          \\ \midrule
\multicolumn{1}{l|}{\multirow{8}{*}{Copilot}} & \multicolumn{1}{c|}{\multirow{4}{*}{GPT4.1}}  & ROCODE-HE      & no                                 & no                                     & no                            & yes                                       & no                                      & no                                       & no                                & no                                              \\ 
\multicolumn{1}{l|}{}                         & \multicolumn{1}{c|}{}                         & ROCODE-MBPP           & no                                 & no                                     & no                            & no                                       & no                                      & no                                       & no                                & no                                              \\
\multicolumn{1}{l|}{}                         & \multicolumn{1}{c|}{}                         & Logparser-AEL         & no                                 & no                                     & no                            & no                                       & no                                      & no                                       & no                                & no                                              \\
\multicolumn{1}{l|}{}                         & \multicolumn{1}{c|}{}                         & Logparser-Drain       & no                                 & no                                     & no                            & no                                       & no                                      & no                                       & no                                & no                                              \\ \cmidrule{2-11} 
\multicolumn{1}{l|}{}                         & \multicolumn{1}{c|}{\multirow{4}{*}{Claude4}} & ROCODE-HE      & yes                                 & yes                                     & partial                            & yes                                       & no                                      & yes                                       & yes                                & no                                              \\
\multicolumn{1}{l|}{}                         & \multicolumn{1}{c|}{}                         & ROCODE-MBPP           & yes                                 & yes                                     & partial                            & yes                                       & no                                      & yes                                       & yes                                & no                                              \\
\multicolumn{1}{l|}{}                         & \multicolumn{1}{c|}{}                         & Logparser-AEL         & no                                 & yes                                     & partial                            & yes                                       & no                                      & yes                                       & yes                                & yes                                              \\
\multicolumn{1}{l|}{}                         & \multicolumn{1}{c|}{}                         & Logparser-Drain       & no                                 & no                                     & no                            & no                                       & no                                      & yes                                       & yes                                & no                                              \\ \bottomrule
\end{tabular}
\end{adjustbox}
\end{table}
\label{RQ1_results}

\textbf{Overall Performance Analysis.}
\Cref{table:RQ1-steps} shows that only in one case (the last column and line 7 in \Cref{table:RQ1-steps}), the multi-agent system successfully reproduces the final results of the studied tasks, indicating that Copilot hardly achieved complete adaptation.
In all cases, the studies LLMs with the multi-agent pipeline of Copilot can initiate the early stages of the workflow. They all perform the identification and reading of files that are necessary (Step 1) and subsequently create or modify the corresponding files (Step 2).
However, when compared with the ground-truth implementations, none of the cases truly complete Step 2, and in most cases, the agents fail to even locate all the essential files in Step 1. 
This suggests that Copilot did not establish a clear understanding of the project structure at the beginning.
As a result, although the requirements of dataset adaption can be inferred from the reference code retained in the repositories, Copilot is unable to generate bug-free code that meets the requirements. Then we allow multi-agent systems to repair the code latter. As Step 5 later shows, one agent ultimately produced the correct final output. This demonstrates the potential of multi-agent systems to complete dataset adaptation tasks.

\textbf{Model-wise Analysis.}
Overall, Claude Sonnet 4 consistently outperforms GPT-4.1 across all steps, achieving higher completion rate and correctness. As shown in \Cref{table:RQ1-steps}, GPT-4.1 performs worse than Claude Sonnet 4 starting from Step 1. GPT-4.1 exhibits limited capability in capturing the overall structure of the repository.
Specifically, although it read the files of the main process, such as \textit{main.py}, it frequently overlooks essential scripts for execution and evaluation (e.g., \textit{run.sh}). 
Also, the task requires the agents to be aware of the dataset structure, either by inspecting example files (e.g., \textit{example\_problem.jsonl}) or by reading the complete dataset contained within compressed archives. GPT-4.1 generally adopts the latter approach but typically executes only a single operation to open the archive and terminates the process upon failure, without performing further exploration or decompression. Consequently, GPT-4.1 conducts only limited exploration of the components within the target dataset.

The performance gap between GPT-4.1 and Claude Sonnet 4 becomes even more evident in Step 2. For files it failed to identify in Step 1, GPT-4.1 naturally makes no modifications, thus omitting several files that actually require changes. Furthermore, its limited understanding of both the dataset and the methodology repository in Step 1 often leads to unreliable adaptations of the target technology. For example, in the ROCODE adaptation task, GPT-4.1 failed to employ the ROCODE framework as specified by the naïve prompt and repository. Instead, it used a vanilla LLM-based code generation process, resulting in an incorrect and incomplete adaptation. As a result, GPT-4.1 fails in nearly all cases in Steps 3, 4, and 5.

In contrast, Claude Sonnet 4 follows a more comprehensive workflow and achieves more faithful implementations in Step 2, largely preserving the structure of the target technology. As shown in \Cref{table:RQ1-steps}, Step 4, in each experiment, Claude Sonnet 4 consistently identifies potential issues in the generated code.
We observe that it often generates additional inputs to test important modules on its own initiative, and even extracts a small portion of the dataset to execute an end-to-end run.
It typically terminates the process only after confirming that the generated code produces plausible outputs. 
However, as indicated in the table, in most cases, Claude Sonnet 4 still fails to reproduce the correct final result. 
Also, in Step 3, sub step 1, we observe that Claude Sonnet 4 consistently generates a command. Although it is not always entirely correct, this behavior is stably present in this step.

\textbf{Step-wise Analysis.}
In Step 1 (reading required files), the multi-agent system consistently identifies the main files that drive the prediction process, such as \textit{main.py}. These files define the primary execution flow of the repository. However, it often misses or misidentifies files involved in other stages of the pipeline, such as evaluation scripts or entry points. In practice, while \textit{README.md} was not explicitly listed among the required files, it proves highly beneficial for the agents, as it typically specifies the entry Bash file and thus reveals the overall execution pipeline. Notably, in the only two cases where Claude Sonnet 4 successfully read all required files  (column 1 line 5 and 6 in
table), the multi-agent selected \textit{README.md} as its first file, suggesting that this file provides crucial contextual guidance for repository comprehension.

Step 2, which involves editing or creating the necessary files, emerges as the primary bottleneck of the pipeline. 
The first sub-step (editing) is closely linked to the outcomes of Step 1: As shown in \cref{table:RQ1-steps}, 2 out of 8 studied cases were successfully completed in Step 1 (column 1, line 5 and 6 in \cref{table:RQ1-steps}), and the multi-agent system also succeeded in the first sub-step of Step 2 for these same two cases. Among the remaining six cases where Step 1 failed, only one managed to identify all the files need to be modified. This pattern aligns with expectations, as a clear understanding of the repository structure increases the likelihood of accurately locating and modifying the relevant files. The second sub-step of Step 2, in contrast, represents the most challenging part of the entire process.  As shown in \Cref{table:RQ1-steps}, none of the cases completely passed this sub-step. In Step 2, one of the challenging tasks can be understanding and leveraging the fields in a dataset. The "field" here refers to a specific data attribute or key (e.g., \textit{prompt}, \textit{canonical\_solution}, or \textit{test} in the HumanEval dataset) that defines part of the dataset’s schema. We observe that the multi-agent system correctly identified field names and assigned them appropriate roles within the data-processing pipeline. For example, the agents correctly identify and use the \textit{prompt} field in HumanEval as the input for generating predictions. However, the adaptation code they produce does not satisfy the requirements for reproducing the original experimental artifacts. For example, in most MBPP cases, Copilot overlooks the fact that one input component of the pipeline requires runnable Python code. However, Copilot places the natural-language problem description there (which should only appear as a comment if included at all). As a result, the generated code cannot reproduce the original experimental artifacts.

In the first sub-step of Step 3, half of the cases the multi-agent produced a clear and valid command for running the full pipeline in in our experiment. In the second sub-step of Step 3, however, all agents failed.
This outcome was unexpected because the naïve prompt explicitly permitted command execution without user confirmation; nevertheless, none of the agents executed the full pipeline on the entire dataset. In several cases the agents executed the pipeline only on a small subset as a smoke test to verify that execution would proceed. We discuss those cases in Step 4.

As we have discussed in the previous paragraph, Step 4 further highlights the performance divergence between models. GPT-4.1 essentially skips this stage, terminating its process after editing the identified files without verifying correctness or debugging potential issues. In contrast, Claude Sonnet 4 exhibits more sophisticated behavior: it routinely pauses after completing each module, designs input cases to test the modified code, reflects on the results, and determines whether further revision is necessary.

In Step 5, as shown in \Cref{table:RQ1-steps}, only one case(the last column and line
7 in \Cref{table:RQ1-steps}) successfully produced the correct final output.
The remaining failures stem from two main causes: incorrect adaptation, such as the misuse of technologies during prediction; and inappropriate testing choices. For example, when adapting ROCODE to HumanEval, the agent incorrectly used the \textit{evaluation.py} file from the HumanEval repository, whereas it should have used the \textit{evalutae\_generated\_code.py} file provided in the ROCODE repository that was intended for dataset adaptation in that context.

\subsubsection{RQ2} \label{RQ2-result}

\textbf{Prompt 1: Providing error messages.} This can resolve failures manifested as a syntax error or runtime error. The error messages in the prompt serve as a guide to help Copilot identify and fix the underlying issues. 
This prompt can also help address runtime errors. By providing the corresponding Python tracebacks with the type and location of an error, this prompt served as a guide to help Copilot identify the underlying issues. 
In our experiments, this prompt can effectively resolve almost all issues that may lead to syntax or runtime errors within three attempts, with the exception of those arising from incorrect indentation.

For example, when checking whether a string ends with a backslash, the code should use two backslashes in the literal to represent one. However, Copilot generated a condition with only one single backslash, which created an incomplete string literal. As a result, Python raised a syntax error, reporting that the parentheses near the backslash were never closed. With Prompt 1, which provides the original error message to the agents, they were able to identify this mistake and correct it by replacing the single backslash with a properly escaped double backslash.

\textbf{Prompt 2: Providing reference code.} 
This type of fixing prompt compensates for the model’s failure to fully read the context in Step 1.
In Step 1, this prompt forced the multi-agent system to read all the required files. In Step 2, sub-step 1, it corrected most of the 15 files missed by the agents (i.e., not created or modified), but 6 extreme cases remained where Copilot failed to create an entire directory; in these cases, additional guidance is required to elicit the correct action of creating a new folder. In sub-step 2, it further reduced the number of incorrectly edited files from 24 to 5. These remaining cases, however, eventually reached a dilemma: after several iterations the agents kept modifying irrelevant code with no impact on the overall logic.
Nevertheless, before reaching this dilemma, Prompt 2 significantly improved the structure of the target repository: with sufficient reference code, the generated implementation better aligns with the ground-truth solution and incorporates most of the key components of the ground-truth solution.
This result also confirms our earlier observation that correct identification of all required files is a key prerequisite for successful implementation (See \Cref{RQ1_results}), highlighting the importance of Step 1 (Reading required files).

Building upon the fixing prompt analysis, we quantified its effectiveness by measuring the similarity of the generated code against the ground truth using JPlag.
JPlag \cite{jplag} is a token-based tool originally designed for plagiarism detection. It measures structural similarity between programs rather than surface-level textual overlap. It compares token sequences extracted from the parse trees of two programs and computes a similarity score based on the proportion of matching code fragments that share equivalent token sequences and structural patterns.
We used \textit{average similarity} from JPlag, which measures the ratio of matched tokens relative to the mean length of the two programs.
For code generated with the naive prompt, the \textit{average similarity} were 7.25\%. After applying the fixing prompt only once, these values increased to 67.14\%, indicating a much closer structural alignment with the ground truth. This demonstrates that the fixing prompt substantially improves the overall structural similarity.

\textbf{Prompt 3: Indicating buggy locations.} 
This prompt directs the agents to reflect on buggy regions, focusing specifically on where the errors occur. By manually pinpointing such locations, it helps mitigate common issues such as the incorrect use of variables or fields. For example, when a variable is misused in the pipeline, the prompt can indicate the exact occurrence that is problematic, thereby providing the model with precise guidance for correction.

In practice, however, the repair ability of Prompt 3 is limited: across our experiments it was applied eight times, and only three resulted in successful corrections. We attribute this to two factors. First, for problems where Prompt 2 is applicable, many simpler errors have already been resolved by it, leaving Prompt 3 to handle relatively more complex cases. 
Second, as discussed in \Cref{design_RQ2}, Prompt 3 is used in cases where no relevant code snippets or functional references are available. In such situations, the task becomes inherently more challenging, as the agents must fully understand both the technology and the dataset, often involving detailed handling of dataset-specific structures. For example, the function name extraction step in the MBPP dataset requires dealing with such special structures.

\section{Related Work}

\noindent \textbf{LLMs in software engineering.}
Large language models (LLMs), have achieved remarkable progress across a wide range of domains in software engineering (SE)\cite{jigsaw, llm4se_survey}. Beyond general-purpose coding assistance, recent research has explored applying LLMs to increasingly specialized SE problems. For example, several works investigate code completion and generation \cite{code_leetcode, ROCODE, repocoder, coding_frank}, while others focus on software testing and automated program repair, where LLMs are used to detect, diagnose, and fix bugs \cite{inferfix, cref, prompt_evaluating, setesting_survey, testing_study}. Additional efforts further examine code translation and refactoring \cite{rustmap, type_migration, refactor}.

\noindent \textbf{LLM-based agents.} Despite the achievements of the function-level code generation paradigm discussed above, LLM-based agents extend these capabilities by iteratively executing code and refining outputs, enabling models to manage more complex, repository-level development workflows. Early examples such as AutoGPT \cite{auto-gpt} and MetaGPT \cite{metagpt} demonstrated that LLM can autonomously plan and perform multi-step tasks. More recent approaches—including SWE-agent \cite{SWE-agent} and RepairAgent \cite{repairagent}—integrate LLMs with compilers, debuggers, and execution environments to iteratively test and improve code. Xia et al. proposed Agentless \cite{agentless}, which employs a three-phase
process that avoids letting LLMs autonomously choose actions or interact with complex tools. Multi-agent frameworks \cite{dynamicllmpoweredagentnetwork, multiagentcollaboration} further simulate team collaboration, dividing software tasks among multiple role-playing agents.

\section{Conclusion}
This study presented the first empirical study of large language model (LLM)–based multi-agent systems for automating dataset adaptation in software engineering. Using GitHub Copilot’s agent mode with GPT-4.1 and Claude Sonnet 4 backends, we assessed their ability to adapt artifacts such as ROCODE and LogHub2.0 across a structured five-step pipeline. Our findings reveal that current multi-agent systems are still unable to generate code with consistent functional equivalence with real-world implementations. Prompt-level interventions provided by human developers offered significant improvements. However, persistent issues in multi-file coding and autonomous debugging indicate that current agents lack robust mechanisms for iterative correction and coordinated planning. These results highlight the limitations of today’s LLM-based multi-agent systems for dataset adaptation. Future research should focus on incorporating execution-aware feedback loops, adaptive prompting, and explicit inter-agent coordination to enhance reliability.

\clearpage

\bibliographystyle{ACM-Reference-Format}
\bibliography{references}

\end{document}